\documentclass[prd,aps,tightenlines,preprintnumbers,amsmath,amssymb,showpacs]{revtex4}
\usepackage{graphicx}
\usepackage{dcolumn}
\usepackage{bm}
\usepackage{hyperref}

\def \beq{\begin{equation}}
\def \eeq{\end{equation}}

\def\lsim{\mathrel{\rlap{\lower4pt\hbox{\hskip1pt$\sim$}}
    \raise1pt\hbox{$<$}}}                
\def\gsim{\mathrel{\rlap{\lower4pt\hbox{\hskip1pt$\sim$}}
    \raise1pt\hbox{$>$}}}                
\newcommand{\bra}[1]{\langle #1|}
\newcommand{\ket}[1]{|#1 \rangle}

\begin{document}

\preprint{EFI 08-27}

\title{Constraining Light Bosons with Radiative $\Upsilon(1S)$ Decays}
\author{David McKeen}
\email{mckeen@theory.uchicago.edu}
\affiliation{Enrico Fermi Institute and Department of Physics, University of Chicago, 5640 South Ellis Avenue, Chicago, IL 60637}
\date{\today}

\begin{abstract}
Light bosons can be found in large classes of theories beyond the standard model.  These light bosons may not be ruled out by current experiments and, indeed, may even provide an explanation for some anomalous experimental results.  The radiative decays of quarkonium ($c \bar c,~b \bar b$) states offer a promising opportunity to investigate such light bosons.  Specifically, we investigate the reach that current CLEO data can have on models with light scalar and pseudoscalar bosons.
\end{abstract}

\maketitle


\begin{section}{Introduction}
Models that contain light CP-even and CP-odd bosons occur in many classes of theories beyond the standard model.  They are a generic feature of models with an expanded Higgs sector.  The extended freedom in the scalar mass matrices allows a wide range of parameter space in which some bosons may be light while reducing their couplings to the $Z$ boson so that LEP would not have observed them.  A particularly attractive example is the next-to-minimal standard supersymmetric model (NMSSM) which solves the $\mu$-problem of the MSSM.  In the NMSSM, light pseudoscalars can arise naturally \cite{Dermisek:2005ar}.  Minimal model-independent light dark matter scenarios also can involve light scalars or pseudoscalars as mediators to serve as an efficient annihilation channel.  Such light bosons should mix with CP-even or -odd Higgs bosons to couple to standard model fermions.  Light scalars and pseudoscalars can also arise generically in models that contain light sgoldstinos \cite{Gorbunov:2000th}.  There are also extra-dimensional gaugephobic models in which the Higgs is decoupled from gauge bosons, allowing it to be light \cite{Cacciapaglia:2006mz, Galloway:2008yh}.
\par
There are hints that light bosons may already have been observed experimentally.  First, if a model contains a pseudoscalar with mass, $m_a<2 m_b$, the LEP limit on the mass of the standard model Higgs boson ($m_h \gsim 114.4~{\rm GeV}$) can be evaded.  This could also provide an explanation for the slight excess of events seen at $m_h\simeq 100~{\rm GeV}$ \cite{Barate:2003sz} since the Higgs' branching ratio to $b$ quarks is reduced \cite{Dermisek:2005gg}.  In addition, these models are also able to accommodate a light dark matter particle that could explain the excess of 511 keV photons from the Galactic center seen by the INTEGRAL satellite \cite{Jean:2003ci,Boehm:2003bt} although recent measurements may cast the explanation in terms of dark matter in some doubt \cite{Weidenspointner:2008zzu}.  The DAMA/LIBRA Collaboration's observations \cite{Bernabei:2008yi} can also be reconciled with null results from CDMS if dark matter is relatively light \cite{Savage:2004fn,Gondolo:2005hh}.  These scenarios are  discussed further in Ref. \cite{Gunion:2005rw}.  In addition, a model containing a pseudoscalar boson with a mass of 214 MeV \cite{He:2005we} can explain the three $\Sigma^+\to p\mu^+\mu^-$ decays of the same dimuon mass seen by the HyperCP Collaboration \cite{Park:2005ek}.  The probability of such a cluster of events in the standard model is regarded as below 1\% \cite{He:2005yn}.  Explanations of this signal as due to a light sgoldstino have also been studied \cite{Gorbunov:2005nu}.  However, recent results from CLEO rule out the explanation of the HyperCP results as being due to a pseudoscalar of mass 214 $\rm MeV$ \cite{:2008hs}.  Furthermore, there are hints that light bosons could be the cause of a 3$\sigma$ deviation from the SM prediction for the process $\pi^0\to e^+ e^-$ at KTeV
 \cite{Dorokhov:2007bd,Kahn:2007ru,Chang:2008np}.
\par
In this paper we investigate the implications of light scalars and pseudoscalars for the decays $\Upsilon (1S)\to\gamma\{\mu^+\mu^-,\pi^+\pi^-,K^+ K^-\}$ and estimate the backgrounds from the radiative return processes $e^+ e^- \to \gamma \gamma^* \to \gamma \{\mu^+\mu^-,\pi^+\pi^-,K^+ K^-\}$.  The radiative return to a muon pair has recently been considered in Ref.\ \cite{Mangano:2007gi}.

The case of light pseudoscalars is presented in Sec. II.  The couplings of the pseudoscalar to standard model fermions and the expression for the signal they generate are given in Sec. IIA.  The background to this process is shown in Sec. IIB.  In Section IIC, the experimental reach with respect to the parameters is discussed.  This is repeated for the case of light scalars in Secs. IIIA, IIIB, and IIIC.  In Sec. IV we conclude.
\end{section}


\begin{section}{Pseudoscalars}
\begin{subsection}{Signal (including scalars)}
We consider a model with a pseudoscalar $A$ and a scalar $S$ coupled to up- and down-type quarks and leptons.  We assume the couplings to be flavor diagonal.  We parameterize the interaction Lagrangian as
\begin{align}
{\cal L}_{A}&=-\left(\frac{iA}{v}\right)\left(g_d m_\ell \bar{\ell} \gamma^5 \ell + g_d m_d \bar{d} \gamma^5 d + g_u m_u \bar{u} \gamma^5 u\right)~~~,
\label{eq:ps_lagrangian}
\\
{\cal L}_{S}&=-\left(\frac{S}{v}\right)\left(\lambda_d m_\ell \bar{\ell} \ell + \lambda_d m_d \bar{d} d + \lambda_u m_u \bar{u} u\right)~~~.
\label{eq:s_lagrangian}
\end{align}
with copies for each family.  For simplicity, we have used equal coupling to leptons and down-type quarks, motivated by grand unified theories.  The results should be straightforwardly generalized to different couplings.  Here, $v=(\sqrt{2}G_F)^{-1/2}\simeq 246\ {\rm GeV}$.  FCNCs are generated by one loop two-quark operators and by four-quark operators.  Limits from flavor changing decays such as $K\to \pi\mu^+\mu^-$ and $B\to K\mu^+\mu^-$ will tend to provide stricter constraints than those from $\Upsilon$ decays.  However, there are examples of models where parameters can be chosen so contributions to these processes can cancel.  In particular, in the NMSSM, there exist ranges of squark mass splittings and chargino masses where such cancellations occur.  For details, see Ref.\ \cite{He:2005we}.  For our purposes we will neglect flavor changing decays to concentrate as an independent probe on the limits that can be derived from $\Upsilon$ decays.  The rate of the radiative decay of the $\Upsilon$ to a pseudoscalar is \cite{Wilczek:1977zn,Nason:1986tr}
\begin{align}
\frac{\Gamma(\Upsilon\to\gamma A)}{\Gamma(\Upsilon\to e^+e^-)}&=\frac{g_d^2 m_b^2 G_F}{\sqrt{2}\pi\alpha}\left(1-\frac{m_A^2}{m_\Upsilon^2}\right){\cal C}_A(x)
\label{eq:ps_decay_rate}
\end{align}
and similarly for a scalar,
\begin{align}
\frac{\Gamma(\Upsilon\to\gamma S)}{\Gamma(\Upsilon\to e^+e^-)}&=\frac{\lambda_d^2 m_b^2 G_F}{\sqrt{2}\pi\alpha}\left(1-\frac{m_S^2}{m_\Upsilon^2}\right){\cal C}_S(x)
\label{eq:s_decay_rate}
\end{align}
Here, $C_F=(N^2-1)/2N=4/3$ for $SU(3)$ and $x=1-m_{A,S}^2/m_\Upsilon^2=2E_\gamma/m_\Upsilon$.  ${\cal C}_A(x)$ and ${\cal C}_S(x)$ describe QCD and relativistic corrections \cite{Nason:1986tr} and are given by
\begin{align}
{\cal C}_{A,S}(x)&=\left[1-\left(\frac{\alpha_s C_F}{\pi}\right)F_{A,S}\left(x\right)\right]~~~.
\label{eq:qcd_corr}
\end{align}
Fig.\ \ref{fig:qcd_corr} shows $F_A(x)$ and $F_S(x)$.  The differential rate is distributed with respect to the photon's center-of-mass angle with the beam as $1+\cos^2\theta_\gamma$.
\begin{figure}
\begin{center}
\rotatebox{270}{\resizebox{60mm}{!}{\includegraphics{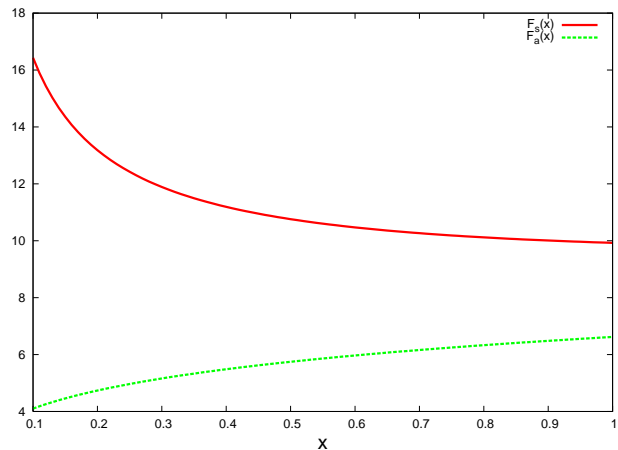}}}
\caption{Plots of $F_A$ and $F_S$ [Eq.\ (\ref{eq:qcd_corr})] as functions of $x=2E_\gamma/m_\Upsilon$.}
\label{fig:qcd_corr}
\end{center}
\end{figure}
\par
For $m_A<2m_\tau$, we take ${\cal B}(A\to\mu^+\mu^-)\simeq 1$.  The rate to two photons is loop suppressed and CP conservation forbids the $s$-wave decay to pairs of pseudoscalar mesons.  Decays with three particles in the final state are taken to be negligible.
\end{subsection}
\begin{subsection}{Background}
The dominant background to the process $\Upsilon(1S)\to\gamma A\to\gamma\mu^+\mu^-$ comes from the radiative return $e^+e^-\to\gamma\gamma^*\to\gamma\mu^+\mu^-$.  The differential cross section is \cite{Mangano:2007gi}
\begin{align}
&\frac{d\sigma}{dm_{\mu\mu}^2}=\left(\frac{\alpha^3}{4\pi \hat{s}}\right) \frac{d\cos\theta_{\gamma}dx d\varphi}{(m_{\mu\mu}^2)(p_{e^-} \cdot k)(p_{e^+} \cdot k)}\Biggl\{(p_{e^-} \cdot p_{\mu^-})^2\Biggr.
\\
\nonumber
&\Biggl.+(p_{e^-} \cdot p_{\mu^+})^2+(p_{e^+} \cdot p_{\mu^-})^2+(p_{e^+} \cdot p_{\mu^+})^2\Biggr.
\\
\nonumber
&\Biggl.+\frac{2m_\mu^2}{m_{\mu\mu}^2}\left[(p_{e^-} \cdot (p_{\mu^-}+p_{\mu^+}))^2+(p_{e^+} \cdot (p_{\mu^-}+p_{\mu^+}))^2\right]\Biggr\}
\end{align}
with $\hat{s}$ the center-of-mass energy squared, $k$ the four momentum of the photon, $\varphi$ the azimuthal angle between the photon and muon, and $x=2 E_{\mu^-} /\sqrt{s}$.  This is plotted in Fig.\ \ref{fig:muon_diffxsection} as a function of dimuon mass for several different limits on the photon's angle.
\begin{figure}
\begin{center}
\rotatebox{270}{\resizebox{60mm}{!}{\includegraphics{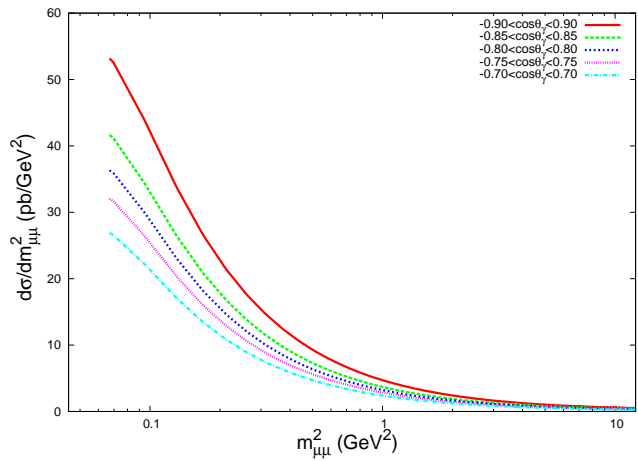}}}
\caption{The cross section for $e^+e^-\to\gamma\mu^+\mu^-$ in the standard model with different cuts on the photon's angle with the beamline.}
\label{fig:muon_diffxsection}
\end{center}
\end{figure}
\end{subsection}
\begin{subsection}{Probing the coupling to fermions}
One can set a lower limit on the value of ${\cal B}(\Upsilon\to A\gamma){\cal B}(A\to\mu^+\mu^-)$  that is observable, or equivalently on $g_d$, by equating the number of events due to the signal to the number due to statistical fluctuations of the background in a certain range of dimuon mass.  We focus on the CLEO experiment, taking an integrated luminosity at the $\Upsilon(1S)$ resonance of $1.06~{\rm fb}^{-1}$ \cite{Besson:2005jv}, $21\times 10^6$ $\Upsilon(1S)$ produced, a resolution of $5~{\rm MeV}$ on the dimuon mass, and a nominal efficiency of $2\%$.  For different values of the pseudoscalar mass we plot the lower bound on the observable value of ${\cal B}(\Upsilon\to\gamma a){\cal B}(A\to\mu^+\mu^-)$ in Fig.\ \ref{fig:muon_pseudo_br_excl}.
\begin{figure}
\begin{center}
\rotatebox{270}{\resizebox{60mm}{!}{\includegraphics{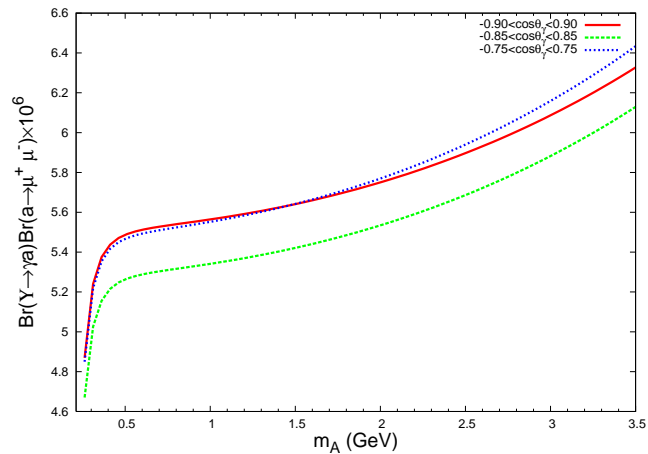}}}
\caption{Lower bound on observable values of ${\cal B}(\Upsilon\to A\gamma){\cal B}(A\to\mu^+\mu^-)$ for a range of pseudoscalar masses with several values of cut on photon angle. It is seen that taking $-0.85 < \cos\theta_\gamma < 0.85$ generates a slightly stronger limit than either of the other two cuts.}
\label{fig:muon_pseudo_br_excl}
\end{center}
\end{figure}
Indeed, CLEO reports limits that are roughly consistent with these estimates in Ref. \cite{:2008hs}, in the range of (several) $10^{-6}$ over the mass range in Fig.\ \ref{fig:muon_pseudo_br_excl}.
\par
If we take ${\cal B}(A\to\mu^+\mu^-)\simeq 1$, this can simply be translated into a limit on the lowest possible value of $g_d$ observable by using Eq.\ (\ref{eq:ps_decay_rate}) to express ${\cal B}(\Upsilon\to\gamma A)$ in terms of $g_d$.  We use $\alpha_s(M_\Upsilon)\simeq 0.2$ and note that, for $m_a,m_s=0$, the QCD corrections given in Eq.\ (\ref{eq:qcd_corr}) are around 40\% and 65\% of the tree-level rates in the pseudoscalar and scalar cases, respectively.  This indicates that there is a substantial amount of theoretical uncertainty and quantities obtained using Eqs.\ (\ref{eq:ps_decay_rate}) and (\ref{eq:s_decay_rate}) could vary quite a bit.  Figure\ \ref{fig:muon_g_excl} shows the limits on $g_d$ with the tree level QCD corrections taken into account.
\begin{figure}
\begin{center}
\rotatebox{270}{\resizebox{60mm}{!}{\includegraphics{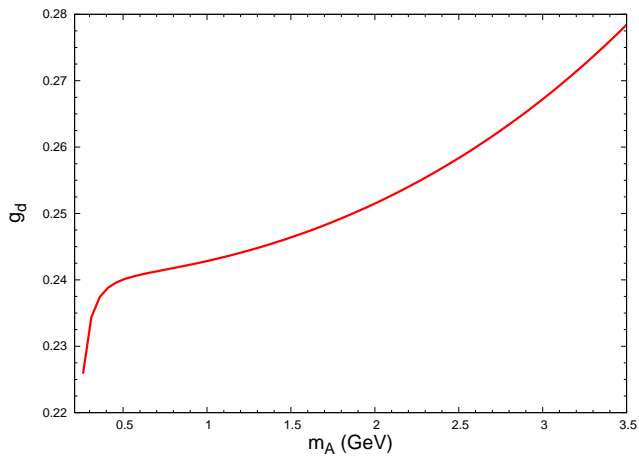}}}
\caption{Lower bound on observable values of $g_d$ for a range of pseudoscalar masses with the photon cut near its optimum: $-0.85 < \cos\theta_\gamma < 0.85$.}
\label{fig:muon_g_excl}
\end{center}
\end{figure}
\par
Figure\ \ref{fig:muon_pseudo_br_excl} shows that there is a slight dependence on the cut on the photon's angle with the beamline and that an optimal value occurs somewhere in the range of $0.75\leq\cos\theta_\gamma\leq 0.85$.  This can be understood as being due to the angular distributions of the signal and background.  The signal has a distribution proportional to $1+\cos^2\theta_\gamma$ while the background is roughly distributed as $(1-\cos^2\theta_\gamma)^{-1}$.  If we denote the cut on $\cos\theta_\gamma$ by $c$, the lower bound on $g_d$ depends on $c$ as
\beq
(g_d^2)_{\rm min}\propto \frac{\left[\int_0^c (1-z^2)^{-1}dz\right]^{1/2}}{\int_0^c (1+z^2)dz}~~~.
\eeq
This function has a maximum at $c \simeq 0.88$.  The background's deviation from the $(1-\cos^2\theta_\gamma)^{-1}$ distribution likely accounts for the minimum occurring at a slightly lower value of $c$.
\end{subsection}
\end{section}


\begin{section}{Scalars}
\begin{subsection}{Signal (additional features)}
We now consider a model with a scalar $S$ coupled to fermions as in Eq.\ \ref{eq:s_lagrangian}.  We define the ratio of the scalar's up- and down-type couplings as $r=\lambda_u/\lambda_d$.  The rate of the radiative decay of the $\Upsilon$ to a scalar is given by Eq.\ \ref{eq:s_decay_rate}.  As mentioned above, the differential rate is distributed as $1+\cos^2\theta_\gamma$.
\par
In the case of the light scalar, determining its branching ratios is slightly more complicated than for a pseudoscalar.  The rate to muons is
\beq
\Gamma(S\to\mu^+\mu^-)=\frac{\lambda_d^2 m_\mu^2 G_F}{4\sqrt{2}\pi}\beta_\mu^3
\eeq
with $\beta_\mu=(1-4m_\mu^2/m_S^2)^{1/2}$.  We estimate the coupling to two pseudoscalar mesons following Refs.\ \cite{Gunion:1989we,Voloshin:1980zf} and references therein.  Heavy quark $(c,b,t)$ loops generate a coupling of the scalar to two gluons
\begin{align}
{\cal L}_{\rm eff}=\frac{\alpha_s}{12\pi}\left(\frac{S}{v}\right)(2\lambda_u+\lambda_d)G_{\mu\nu}^a G^{a\mu\nu}~~~.
\end{align}
The conformal anomaly then relates this term to the trace of the energy momentum tensor
\beq
\Theta_\mu^\mu=-\frac{b\alpha_s}{8\pi}G_{\mu\nu}^a G^{a\mu\nu}+\sum_{i=u,d,s} m_i \bar{\psi}_i \psi_i~~~,
\eeq
where $b=11-2n_F/3$ is the coefficient of the QCD $\beta$ function, $\beta(g_s)=-bg_s^3/16\pi^2$.  Putting this all together one obtains the interaction Lagrangian
\begin{align}
{\cal L}_{\rm int}&=\frac{S}{v}\left\{\frac{2(2\lambda_u+\lambda_d)}{27}\left(\sum_{i=u,d,s}  m_i \bar{\psi}_i \psi_i-\Theta_\mu^\mu\right)\right.~~~.
\\
\nonumber
&\quad\quad\left.-\sum_{i=u,d,s} \lambda_i m_i \bar{\psi}_i \psi_i \right\}~~~.
\end{align}
One can then use a chiral Lagrangian and make the replacement
\begin{align}
\sum_{i=u,d,s} m_i \bar{\psi}_i \psi_i\to \frac{f_\pi^2 m_0}{8}{\rm Tr}\left(\Sigma^\dagger M\right)+{\rm h.c.}
\end{align}
with  $\Sigma$ a matrix containing the pseudoscalar octet, $\Sigma=\exp[(2i/f_\pi)\lambda^a \pi^a]$, $m_0$ related to the quark condensate as $\langle\bar{q}_i q_i\rangle \simeq f_\pi^2 m_0$, and $M$ a matrix containing the quark masses, $M={\rm diag}(m_u,m_d,m_s)$.  Similarly we replace
\begin{align}
\sum_{i=u,d,s} \lambda_i m_i \bar{\psi}_i \psi_i\to \frac{f_\pi^2 m_0}{8}{\rm Tr}\left(\Sigma^\dagger M^\prime\right)+{\rm h.c.}
\end{align}
with $M^\prime={\rm diag}(\lambda_u m_u,\lambda_d m_d,\lambda_d m_s)$.  We write the trace of the energy momentum tensor as
\begin{align}
\Theta_\mu^\mu=-2\partial^\mu \pi^+ \partial_\mu \pi^-+4m_\pi^2 \pi^+\pi^-+\dots~~~.
\label{eq:enmomtrace}
\end{align}
We then find
\begin{align}
\bra{\pi^+\pi^-}\sum_{i=u,d,s} m_i \bar{\psi}_i \psi_i\ket{0}&=m_0(m_u+m_d)\nonumber
\\
&\simeq m_{\pi^{\pm}}^2~~~,
\end{align}
\begin{align}
&\bra{\pi^+\pi^-}\sum_{i=u,d,s}  \lambda_i m_i \bar{\psi}_i \psi_i\ket{0}=\lambda_d m_0(rm_u+m_d)\nonumber
\\
&\quad\quad\simeq\frac{\lambda_d}{2}\left[\left(1+r\right)m_{\pi^{\pm}}^2+\left(1-r\right)\Delta m_{K}^2\right]
\end{align}
where $\Delta m_{K}^2=m_{K^{0}}^2-m_{K^{\pm}}^2$.  (Electromagnetic shifts of the pseudoscalar masses have been ignored for simplicity.)  Using Eq. (\ref{eq:enmomtrace}) one can evaluate the matrix element involving the trace of the energy momentum tensor as
\begin{align}
&\bra{\pi^+\pi^-}\Theta_\mu^\mu\ket{0}\simeq q^2+2m_{\pi^{\pm}}^2=m_S^2+2m_{\pi^{\pm}}^2~~~.
\end{align}
This is used to find
\begin{align}
\frac{\Gamma(S\to\mu^+\mu^-)}{\Gamma(S\to\pi\pi)}&\simeq\frac{243 m_\mu^2}{m_S^2}\left\{1+2r+\left(\frac{31+35r}{4}\right)\frac{m_\pi^2}{m_S^2}\right.
\label{eq:chiral_br} 
\\
\nonumber
&\quad\quad\left.+\frac{27}{4}\left(1-r\right)\frac{\Delta m_K^2}{m_S^2}\right\}^{-2}\frac{\beta_\mu^3}{\beta_\pi}
\end{align}
with $\beta_\pi=(1-4m_\pi^2/m_S^2)^{1/2}$ and both charged and neutral pions included.  If we set $r=1$ we find
\begin{align}
\frac{\Gamma(S\to\mu^+\mu^-)}{\Gamma(S\to\pi\pi)}\simeq\frac{27 m_\mu^2}{m_S^2}\left\{1+\left(\frac{11}{2}\right)\frac{m_\pi^2}{m_S^2}\right\}^{-2}\frac{\beta_\mu^3}{\beta_\pi}
\end{align}
which agrees with Eq.\ (2.50) of \cite{Gunion:1989we}.  However, above kaon threshold the chiral Lagrangian picture may no longer hold.  For $m_S>2m_K$ we follow Ref. \cite{Gunion:1989we} and use a perturbative spectator model to estimate the scalar's partial widths.  Denoting $\Gamma(s\to\mu^+\mu^-)$ as $\Gamma_{\mu\mu}$, etc., we take
\begin{align}
&\Gamma_{\mu\mu}:\Gamma_{u\bar{u},d\bar{d}}:\Gamma_{s\bar{s}}:\Gamma_{gg}:\Gamma_{NN}=\lambda_d^2 m_\mu^2\beta_\mu^3:3\lambda_{u,d}^2m_{u,d}^2\beta_\pi^3
\label{eq:pert_spec_br}
\\
&:3\lambda_d^2 m_s^2\beta_K^3:\left(\frac{\alpha}{\pi}\right)^2 \frac{m_s^2}{3} \left\{3(\lambda_u^2+\lambda_d^2)-(\lambda_u^2+\lambda_d^2)\beta_\pi^3-\lambda_d^2\beta_K^3\right\}:3\lambda_{u,d}^2m_{u,d}^2\beta_N^3\nonumber
\end{align}
where $N$ denotes a nucleon.  We equate the amplitude for pion production to that of nucleon production, ignoring differences between diquark and quark pair creation.  As in \cite{Gunion:1989we}, we determine $m_u$, $m_d$, $m_s$, and $\alpha_s$ by matching the chiral model onto the perturbative spectator model in their overlapping regions of applicability.  We find that $m_u=m_d=50~{\rm MeV}$, $m_s=450~{\rm MeV}$ and $\alpha_s=0.47$ allow the perturbative spectator model to reproduce the chiral model fairly well.  Current quark masses do not allow for as good agreement between the two models.  We ignore the $u$ and $d$ quark content in the kaons and the $\eta$ and approximate $\Gamma_{\pi^+\pi^-}+\Gamma_{\pi^0\pi^0}\simeq\Gamma_{u\bar{u}}+\Gamma_{d\bar{d}}$, $\Gamma_{K^+K^-}+\Gamma_{K^0K^0}+\Gamma_{\eta\eta}\simeq\Gamma_{s\bar{s}}$.  We use flavor $SU(3)$ to estimate $\Gamma_{\pi^+\pi^-}\simeq 2\Gamma_{\pi^0\pi^0}$ and $\Gamma_{K^+K^-}\simeq\Gamma_{K^0K^0}\simeq (9/8)\Gamma_{\eta\eta}$ above $\eta$ threshold.  Note that we have used the wavefunction $\eta=(\bar{u}u+\bar{d}d+2\bar{s}{s})/\sqrt{6}$.  Using the more phenomenologically motivated $\eta\simeq(\bar{u}u+\bar{d}d+\bar{s}{s})/\sqrt{3}$ will decrease the importance of the $\eta\eta$ channel relative to kaons.  For simplicity, we ignore $\pi^0$-$\eta$ production, which vanishes in the limit $\lambda_u m_u=\lambda_d m_d$.  Higher multiplicity final states should not enter into the picture until $\rho$ pair production threshold is crossed.  We assume that these final states are phase-space suppressed.  In this approximation, we find that the scalar's branching ratio to baryons is always less than 3\%.  We include baryon production in our determination of other branching ratios but do not plot it for clarity.  Further studies with these assumptions relaxed are warranted.  We plot the estimates of the branching ratios of the scalar to muons, pions and kaons as a function of $m_S$ for several values of $r$ in Figs.\ \ref{fig:scalar_br_muon}-\ref{fig:scalar_br_total}.
\begin{figure}
\begin{center}
\rotatebox{270}{\resizebox{60mm}{!}{\includegraphics{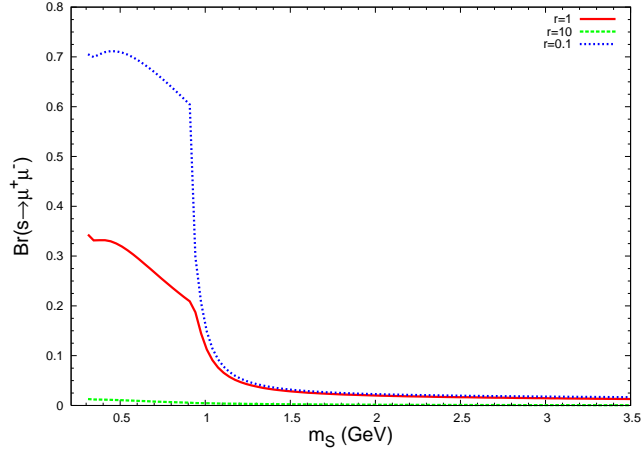}}}
\caption{${\cal B}(S\to\mu^+\mu^-)$ as a function of $m_S$ for several values of $r=\lambda_u/\lambda_d$, obtained from Eqs.\ (\ref{eq:chiral_br}) and (\ref{eq:pert_spec_br}).}
\label{fig:scalar_br_muon}
\end{center}
\end{figure}
\begin{figure}
\begin{center}
\rotatebox{270}{\resizebox{60mm}{!}{\includegraphics{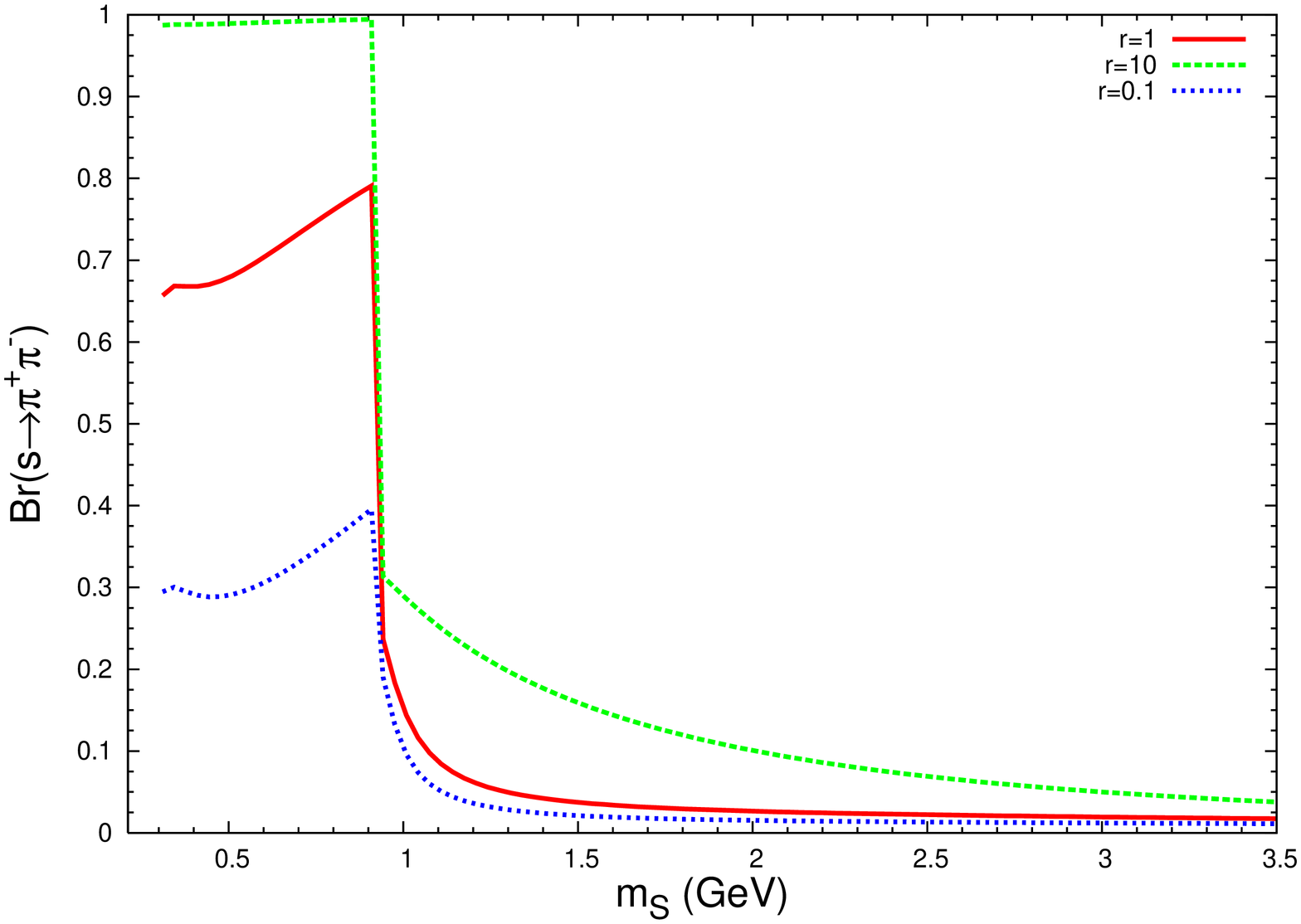}}}
\caption{${\cal B}(S\to\pi^+\pi^-)$ obtained from Eqs.\ (\ref{eq:chiral_br}) and (\ref{eq:pert_spec_br}).}
\label{fig:scalar_br_pion}
\end{center}
\end{figure}
\begin{figure}
\begin{center}
\rotatebox{270}{\resizebox{60mm}{!}{\includegraphics{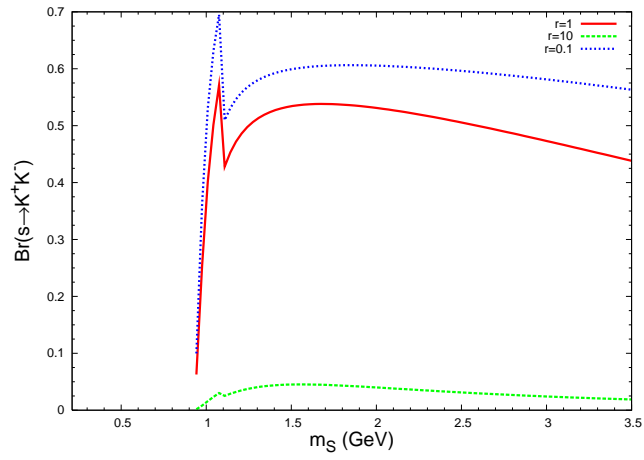}}}
\caption{${\cal B}(S\to K^+K^-)$ obtained from Eqs.\ (\ref{eq:chiral_br}) and (\ref{eq:pert_spec_br}).  The jump at $m_S\simeq 1.1 {\rm GeV}$ is due to $\eta\eta$ production.}
\label{fig:scalar_br_kaon}
\end{center}
\end{figure}
\begin{figure}
\begin{center}
\rotatebox{270}{\resizebox{60mm}{!}{\includegraphics{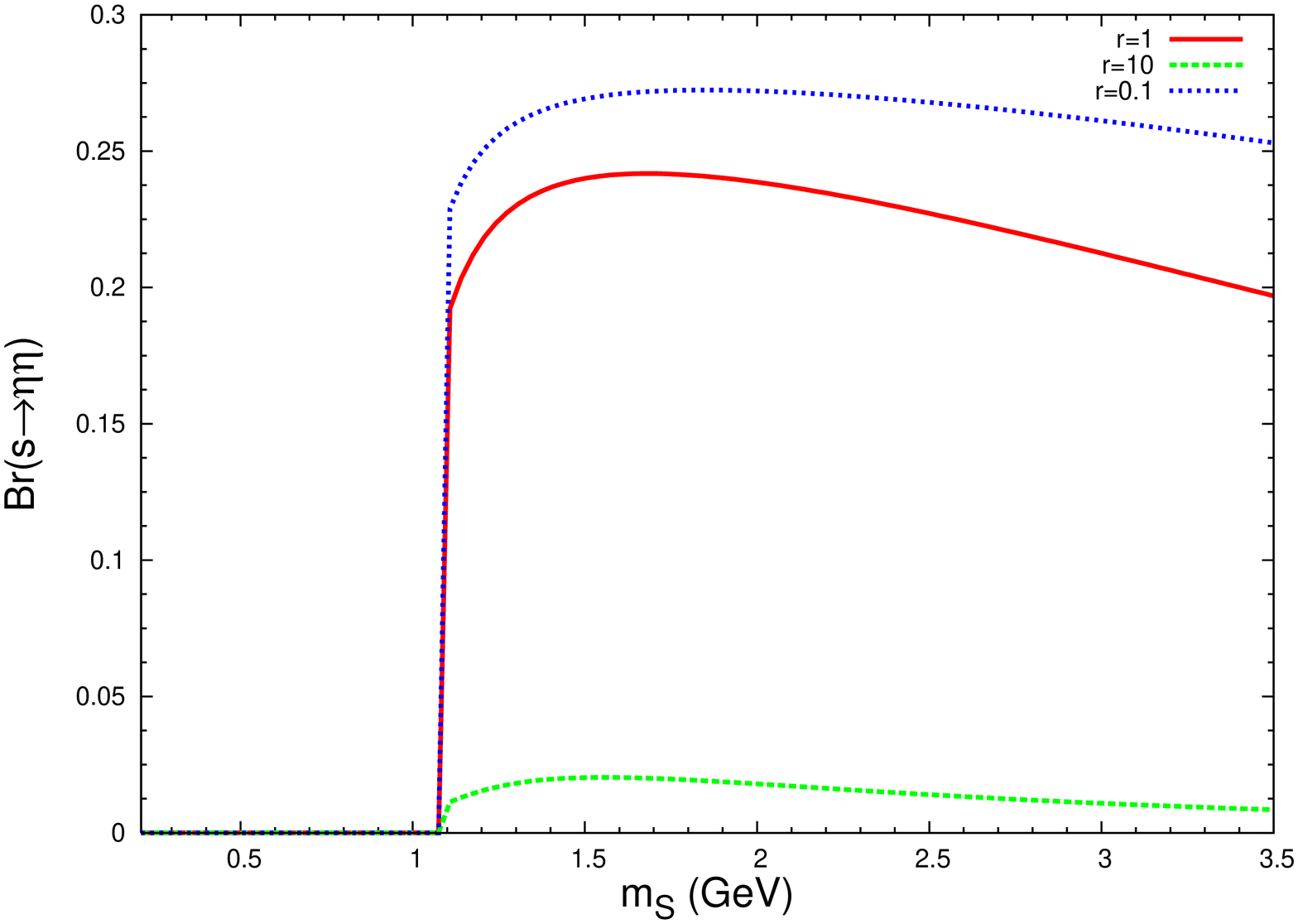}}}
\caption{${\cal B}(s\to \eta\eta)$ obtained from Eqs.\ (\ref{eq:chiral_br}) and (\ref{eq:pert_spec_br}).}
\label{fig:scalar_br_eta}
\end{center}
\end{figure}
\begin{figure}
\begin{center}
\rotatebox{270}{\resizebox{60mm}{!}{\includegraphics{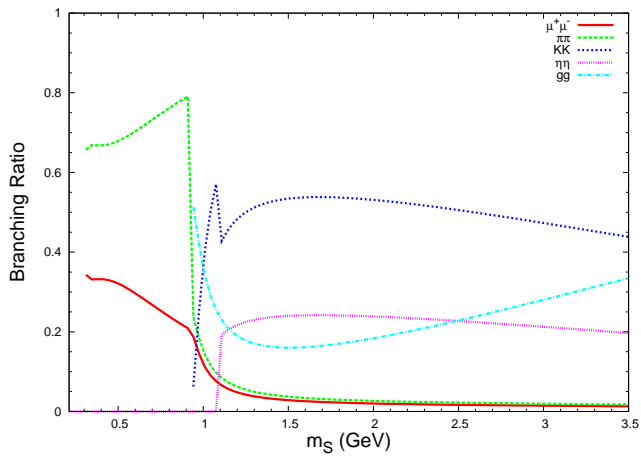}}}
\caption{Scalar's branching ratios as a function of $m_S$ for $r=\lambda_u/\lambda_d=1$, obtained from Eqs.\ (\ref{eq:chiral_br}) and (\ref{eq:pert_spec_br}).  Baryon production is included in this calculation but is found to be less than three percent for all values of $m_S$ and is not plotted for clarity.}
\label{fig:scalar_br_total}
\end{center}
\end{figure}
\end{subsection}
\begin{subsection}{Background}
We take as the dominant background for $\Upsilon(1S)\to\gamma S\to\gamma \pi^+ \pi^-$ the radiative return process $e^+e^-\to\gamma\gamma^*\to\gamma\pi^+\pi^-$ and similarly for the background to $\Upsilon(1S)\to\gamma S\to\gamma K^+ K^-$.  We use vector meson dominance to write the pion and kaon electromagnetic form factors as
\beq
F_{\pi,K}(Q^2)=\sum_{V=\rho,\omega,\phi}N_{V\{\pi,K\}}\left(\frac{m_V^2}{m_V^2-Q^2+i m_V \Gamma_V}\right)~~~,
\eeq
with $N_{\rho \pi}=1$, $N_{\omega  \pi}=N_{\phi \pi}=0$, $N_{\rho K}=1/2$, $N_{\omega K}=1/6$, and $N_{\phi K}=1/3$.  The differential cross sections, $d\sigma/dm_{\pi\pi}$ and $d\sigma/dm_{KK}$, are plotted in Figs.\ \ref{fig:pion_diffxsection} and \ref{fig:kaon_diffxsection}.
\begin{figure}
\begin{center}
\rotatebox{270}{\resizebox{60mm}{!}{\includegraphics{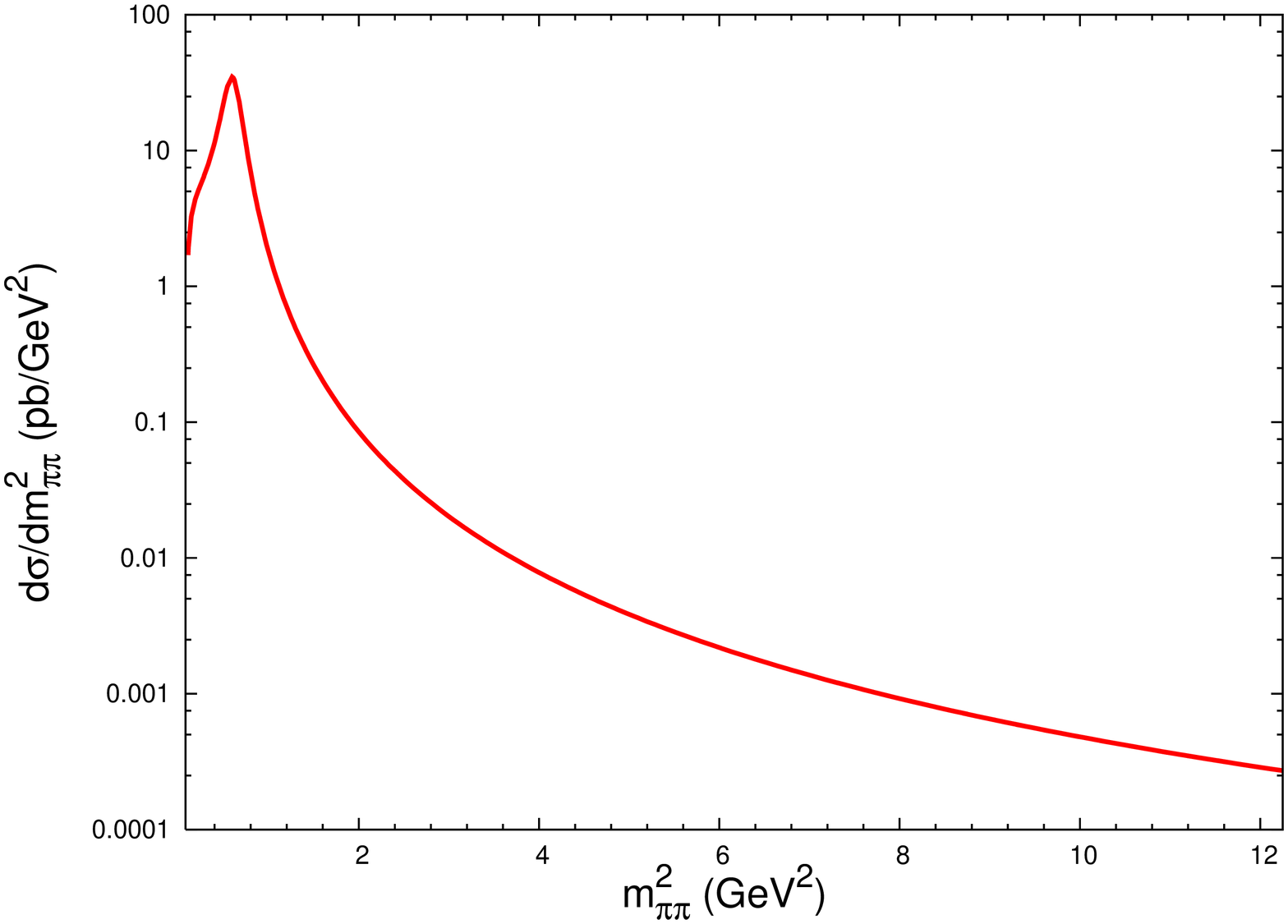}}}
\caption{The cross section for $e^+e^-\to\gamma\gamma^*\to\gamma\pi^+\pi^-$ with $-0.85 < \cos\theta_\gamma < 0.85$.}
\label{fig:pion_diffxsection}
\end{center}
\end{figure}
\begin{figure}
\begin{center}
\rotatebox{270}{\resizebox{60mm}{!}{\includegraphics{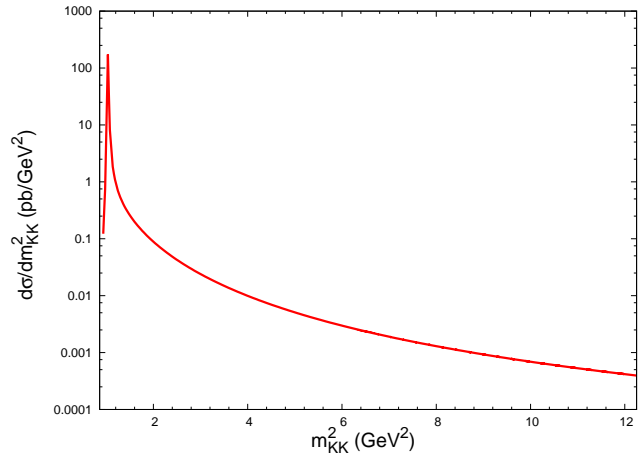}}}
\caption{The cross section for $e^+e^-\to\gamma\gamma^*\to\gamma K^+K^-$ with $-0.85 < \cos\theta_\gamma < 0.85$.}
\label{fig:kaon_diffxsection}
\end{center}
\end{figure}
\end{subsection}
\begin{subsection}{Probing the coupling to fermions}
We can perform a similar analysis of the radiative return background to pion and kaon pairs to set a lower limit on the observable values of ${\cal B}(\Upsilon\to\gamma S){\cal B}(S\to \pi^+ \pi^-)$ and ${\cal B}(\Upsilon\to\gamma S){\cal B}(S\to K^+ K^-)$.  These are plotted in Figs.\ \ref{fig:pion_br_excl} and \ref{fig:kaon_br_excl}.
\begin{figure}
\begin{center}
\rotatebox{270}{\resizebox{60mm}{!}{\includegraphics{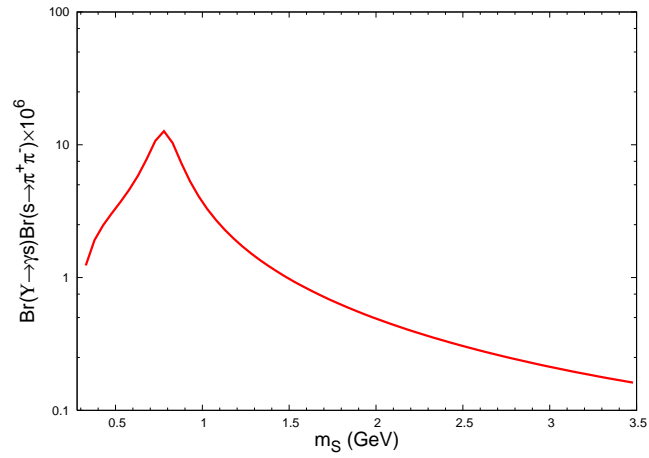}}}
\caption{Lower bound on observable values of ${\cal B}(\Upsilon\to S\gamma){\cal B}(S\to \pi^+ \pi^-)$ for a range of scalar masses with $-0.85 < \cos\theta_\gamma < 0.85$.}
\label{fig:pion_br_excl}
\end{center}
\end{figure}
\begin{figure}
\begin{center}
\rotatebox{270}{\resizebox{60mm}{!}{\includegraphics{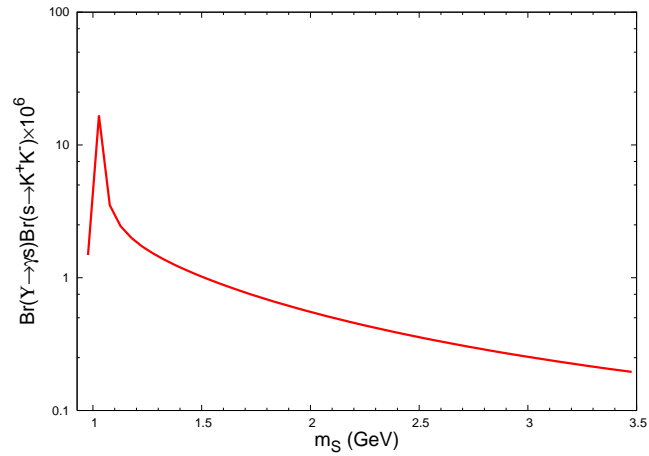}}}
\caption{Lower bound on observable values of ${\cal B}(\Upsilon\to S\gamma){\cal B}(S\to K^+ K^-)$ for a range of scalar masses with $-0.85 < \cos\theta_\gamma < 0.85$.}
\label{fig:kaon_br_excl}
\end{center}
\end{figure}
We can then use Eq.\ (\ref{eq:s_decay_rate}) with the branching ratios of $S$ obtained from Eqs.\ (\ref{eq:chiral_br}) and (\ref{eq:pert_spec_br}) to translate the branching ratio limits to limits on observable values of $\lambda_d$.  We use the same naive efficiency and resolution as in the pseudoscalar analysis.  The limits that come from $\gamma\mu^+\mu^-$, $\gamma\pi^+\pi^-$, and $\gamma K^+K^-$ final states will depend on the value of $r=\lambda_u/\lambda_d$.  These limits, for several values of $r$, are plotted in Figs.\ \ref{fig:lambda_limit_1}-\ref{fig:lambda_limit_10}.
\begin{figure}
\begin{center}
\rotatebox{270}{\resizebox{60mm}{!}{\includegraphics{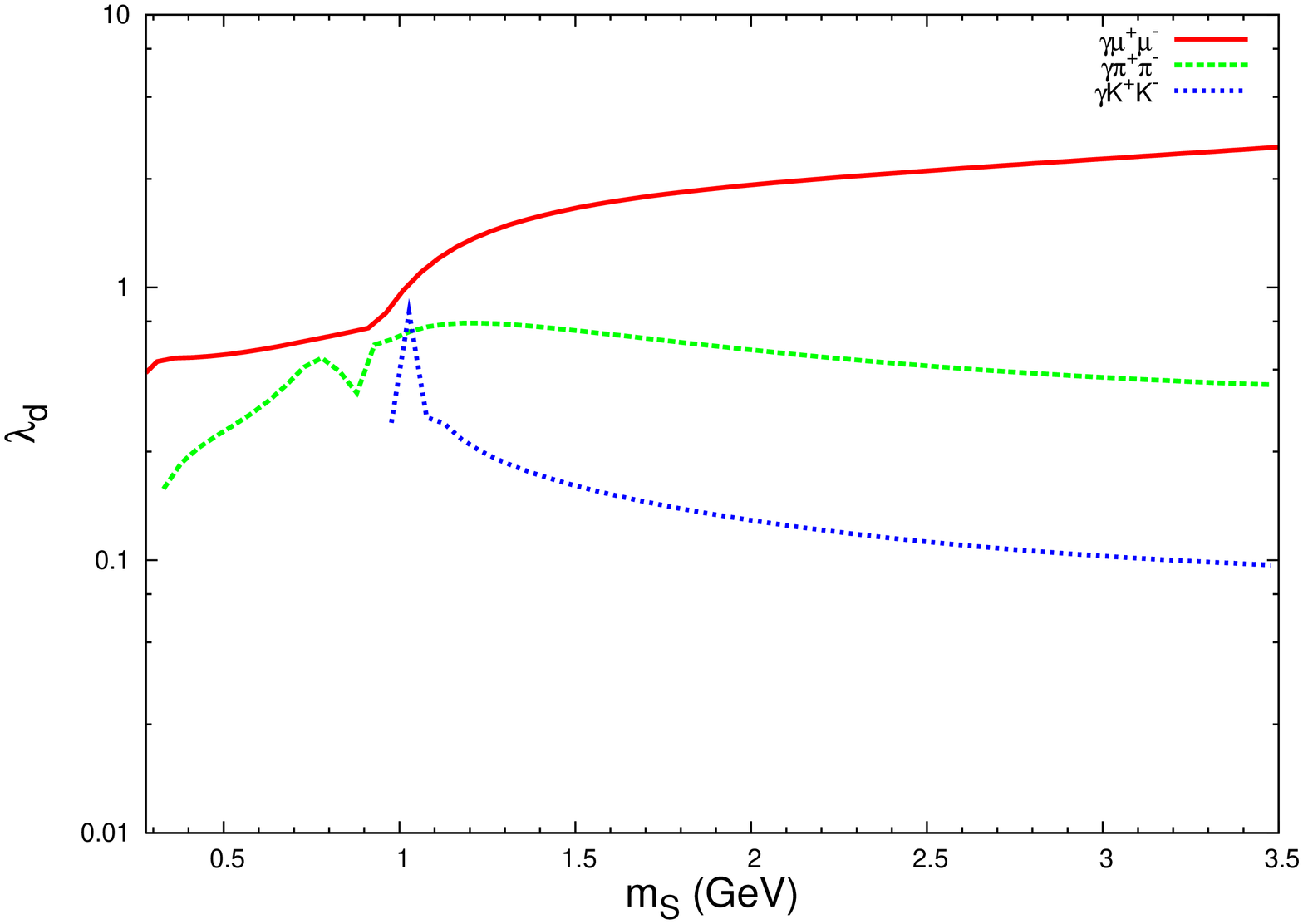}}}
\caption{Lower bound on observable values of $\lambda_d$ obtained from the final states indicated with $-0.85 < \cos\theta_\gamma < 0.85$ for $r=1$.}
\label{fig:lambda_limit_1}
\end{center}
\end{figure}
\begin{figure}
\begin{center}
\rotatebox{270}{\resizebox{60mm}{!}{\includegraphics{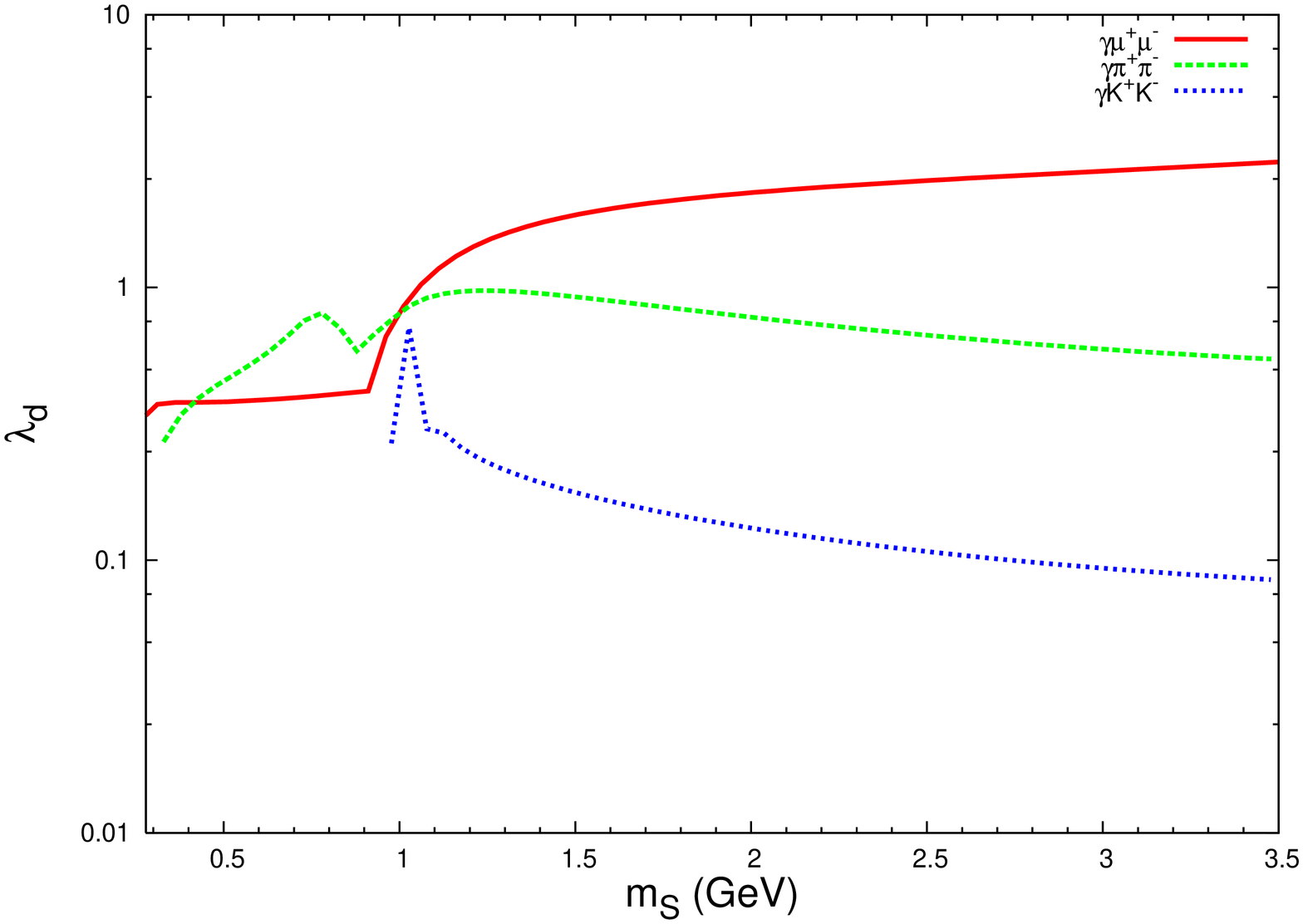}}}
\caption{Lower bound on observable values of $\lambda_d$ for $r=0.1$.}
\label{fig:lambda_limit_01}
\end{center}
\end{figure}
\begin{figure}
\begin{center}
\rotatebox{270}{\resizebox{60mm}{!}{\includegraphics{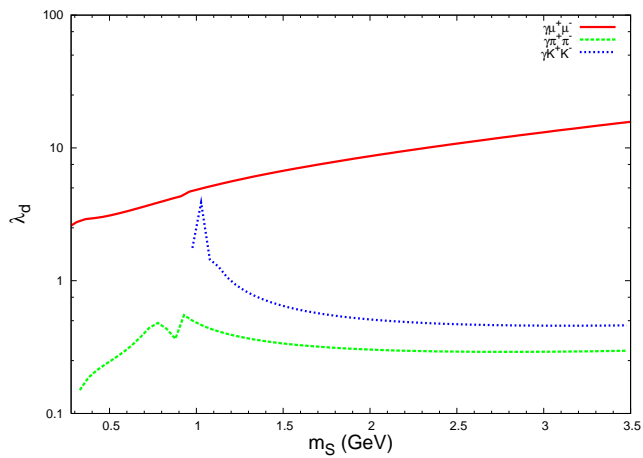}}}
\caption{Lower bound on observable values of $\lambda_d$ for $r=10$.}
\label{fig:lambda_limit_10}
\end{center}
\end{figure}
We observe that the limit due to the pion final state becomes more stringent as $r$ is increased.  For $r>1$ the dimuon final state does not offer very useful bounds and the hadronic final states become more important.
\par
For large $m_{\pi\pi}$ or $m_{KK}$, the background is suppressed since the electromagnetic form factors are far from resonance.  Thus, we see that we can probe down to smaller couplings than in the dimuon case where the increased resolution becomes a limiting factor.  At large enough $m_{\pi\pi}$ or $m_{KK}$, we do not expect any background.  Requiring that we see three signal events limits $\lambda_d\gsim 0.2$.  This can be thought of as an absolute lower bound on observable values of the coupling.
\par
Assuming that the dominant background is due to the radiative return process, the minimum coupling that can be probed in both the scalar and pseudoscalar cases scales as
\begin{align}
\left(\frac{1}{N}\sqrt{\frac{\sigma \int{\cal L}~dt}{\epsilon}}\right)^{\frac{1}{2}}
\end{align}
where $N$ is the number of quarkonia produced, $\sigma$ is the cross section of the radiative return background, $\int{\cal L}~dt$ is the integrated luminosity, and $\epsilon$ is the efficiency.  At a given $q^2$, $\sigma\sim \hat{s}^{-1}=M_V^{-2}$, where $M_V$ is the mass of the resonance.  Then the minimum coupling scales as $(\epsilon M_V^2 N^2/\int{\cal L}~dt)^{-1/4}$.
\end{subsection}
\end{section}


\begin{section}{Conclusions}
Radiative decays of the $\Upsilon(1S)$ can provide useful information about light scalars and pseudoscalars.  There is solid motivation for considering light spinless particles.  A model-independent extension of the scalar sector can generically include light scalars or pseudoscalars that would have escaped detection due to the increased freedom in the couplings of the scalar sector.  Such models can accommodate light dark matter that may be present.  Such scalars also occur in the NMSSM, models with light sgoldstinos, and ``gaugephobic" Higgs scenarios.  In addition, light spinless particles may help to shed light on several anomalous experimental observations.  Quarkonium offers an excellent environment to probe the tree level coupling of light scalars and pseudoscalars to fermions and provides results that are complementary to those that involve flavor changing couplings.  These results can help to tease out the nature of a scalar sector beyond the standard model.
\par
We have seen that current CLEO data on the $\Upsilon(1S)$ resonance can limit couplings down to $g_d,\lambda_d \gsim 0.25$ in the Lagrangians of Eqs.\ (\ref{eq:ps_lagrangian}) and (\ref{eq:s_lagrangian}).  In particular, they are able to rule out \cite{:2008hs} the interpretation of the HyperCP signal as due to a light pseudoscalar Higgs which requires $g_d\simeq0.4$ \cite{He:2005we}.  Looking at Fig.\ \ref{fig:lambda_limit_1}, current CLEO data can also directly rule out a standard model Higgs boson ($\lambda_d=1$, $r=1$) up to masses of about 1 GeV in the dimuon channel.  This is subject to some uncertainty in the branching ratios, although chiral perturbation theory should not be suspect in this area.
\par
Other models can also be constrained.  In models with light sgoldstinos the coupling in the interaction Lagrangian of Eq.\ \ref{eq:ps_lagrangian} is given by $g/v=A_f/F$ and in Eq.\ \ref{eq:s_lagrangian} by $\lambda/v=A_f/F$ where $\sqrt{F}$ is the scale of SUSY breaking and $A_f$ is a soft trilinear coupling.  Limiting $g_d,\lambda_d \gsim 0.25$ constrains $F/A_f\gsim 1\ {\rm TeV}$.  If one sets $A_f\simeq 100\ {\rm GeV}$ then one can limit $\sqrt{F}\gsim 315\ {\rm GeV}$.
\par
In gaugephobic Models, one typically finds $\lambda_d\lsim 0.1$ \cite{Galloway:2008yh}.  Current data are seen above to approach this limit.  Further study would be welcome.
\par
Stated more generally, one can imagine the interactions of Eqs.\ (\ref{eq:ps_lagrangian}) and (\ref{eq:s_lagrangian}) suppressed by the scales $\Lambda_A=v/g_d$, $\Lambda_S=v/\lambda_d$ which are probed up to  $\Lambda_{A,S}\simeq 1\ {\rm TeV}$.  The scale that can be reached at another resonance should go as $\Lambda_{A,S}\sim(\epsilon M_V^2 N^2/\int{\cal L}~dt)^{1/4}$.  The theoretical uncertainty involved in obtaining these results is fairly large, however, and findings could differ somewhat from those presented here.  In particular, a more rigorous incorporation of hadronic decays could help shore up conclusions that can be drawn about light spinless particles coupled to fermions.
\end{section}


\begin{acknowledgments}
The author would like to thank J. L. Rosner for numerous discussions and suggestions.  T. Skwarnicki and N. Sultana also contributed helpful comments on early drafts.  This work was supported in part by the United States Department of Energy under Grant No. DE-FG02-90ER40560.
\end{acknowledgments}


\end{document}